\documentclass[]{article}

\begin{document}
\newcommand{\bra}[1]{{}_{{}_#1}\!\!\!\!<\!\!0|}
\newcommand{\ket}[1]{|0\!\!>\!\!\!{}_{{}_#1}}
\newcommand{\braa}{<\!\!0|}
\newcommand{\kett}{|0\!\!>}
\newcommand{\ze}{^{(0)}}
\title{A Free-Algebraic Solution for the Planar Approximation}
\author{O. Haan \thanks{email: ohaan@gwdg.de}\\Gesellschaft f\"ur wissenschaftliche Datenverarbeitung mbH \\ G\"ottingen}
\date{}
\maketitle
\begin{abstract}
An explicit solution for the generating functional of n-point functions in the planar
approximation is given in terms of two sets of free-algebraic annihilation and creation operators.
\end{abstract}

\section{Introduction}
Thirty years ago, the 1/N expansion \cite{thooft74} was introduced as a new scheme for a perturbative treatment of QCD. The zeroth order of this expansion - the large N limit - includes all planar Feynman diagrams of the theory. This "planar approximation" was argued to give a qualitative correct description of strong interactions \cite{witten79}. The solution for the planar approximation was sought for in terms of a "master field", from which the vacuum expecation values of all $U(N)$ invariant expressions should be calculable \cite{witten80}. For zero-dimensional theories involving one matrix-valued field, the master field has been found \cite{brezin78}. For higher dimensional theories it was argued in \cite{itzykson80,haan81}, that the master field could not be classical in the sense of a single saturating configuration for the path integral defining the theory.

It was realised independently by several authors, that the master field for the planar sector of field theory should be formulated in a framework, which is known in mathematics as "free algebra". The relevant setting here is the representation of a generalised Cuntz algebra \cite{cuntz77} in a free Fock space of states, which are created by creation operators $a_i{}\!^+$ from a vacuum state annihilated by $a_i$, the adjoint of $a_i{}\!^+$:
\begin{eqnarray}
|i_1,i_2,\ldots,i_n> & = & a_{i_1}{}\!\!\!^+ a_{i_2}{}\!\!\!^+ \cdots a_{i_k}{}\!\!\!^+ \kett\\a_i \kett & = & 0.
\end{eqnarray}
The term "free" is used in the mathematical sense and expresses the fact that the product of two free-algebraic creation operators does not obey any constraint (it is not commuting nor anticommuting). Instead the products of creation and annihilation operators $a_i{}\!^+$ and $a_i$ obey the free-algebraic relations
\begin{equation}
\label{prod}
 a_i a_j{}\!^+ = \delta_{ij}\:,\;\;\sum_{i}a_i{}\!^+a_i = 1 - \kett\braa. 
\end{equation}
 
In 1980 I used these operators to define "planar fields" for the non-interacting field theory \cite{haan80}, Halpern \cite{halpern81a} and Halpern and Schwartz \cite{halpern81} introduced them in a Hamiltonian approach  and Cvitanovic et. al.  used "non-commuting sources" with similar properties to define generating functionals for various n-point functions in the planar approximation \cite{cvitanovic82}. 

More than ten years later the work on free random variables by Voiculescu et al. \cite{voiculescu92} and the suggestion of Singer to represent the large N master field in the context of free groups \cite{singer95} stimulated new studies of the large N limit. Douglas \cite{douglas94} and independently Gopakumar and Gross \cite{gopakumar95} found the following formal construction of the master field $\hat{M_l}$ for interacting theories in terms of creation and annihilation operators of the free Fock space as defined in (1-3):

\begin{eqnarray}
	\hat{M_i}=a_i + \sum_n \sum_{i_1\cdots i_n} x_{i,i_1,\cdots,i_n}a_{i_n}{}\!\!\!^+\cdots a_{i_1}{}\!\!\!^+.
\end{eqnarray}
Unfortunately the construction of this master field involves the connected n-point functions $x_{i_1,\cdots,i_n}$ of the theory, which are just the objects one would like to calculate. 

In two papers \cite{halpern99a,halpern99b} Halpern and Schwartz constructed a master field by generalising the algebra (1-3) to the case of interacting fields. From this new algebra they derived an infinite set of polynomials in the master fields with vanishing vacuum expectation values. These relations for the first time gave a set of linear equations for the n-point functions of the large N theory. This set of equations can be generated from the functional 
\begin{eqnarray}
\label{hs}
\braa(1-\beta_lG_l(\phi) + \beta_l\beta_{l'}E_{ll'}(\phi))^{-1}\kett = 1,
\end{eqnarray}
involving a non-commuting source $\beta_l$. For a field theory with cubic self interaction $G_l$ and $E_{ll'}$ are given in terms of the fields $\phi_l$, the kinetic operator $\Delta_{ll'}$ and the one-point function $z_l$ as:
\begin{eqnarray}
	G_l(\phi)= \Delta_{ll'}\phi_{l'} + g\phi_l{}^2, \hspace{1em}E_{ll'}(\phi)=\Delta_{ll'}+g\delta_{ll'}(\phi_l+z_l).
\end{eqnarray}

In this paper I will derive an explicit representation of the n-point functions $z_{i_1\cdots i_n}$ in terms of two independent sets of free-algebraic operators $a_l,a_l{}\!^+$ and  $b_l,b_l{}\!^+$ and their corresponding vacua $\ket{a}, \ket{b}$, which for the case of cubic self interaction reads:

\begin{eqnarray}
\label{nrep1}
	z_{i_1,\cdots,i_n}=\bra{a}\bra{b}a_{i_1}\cdots a_{i_n}(1-S(a,a^+,b,b^+))^{-1}\ket{b}\ket{a}
\end{eqnarray}
with
	\begin{eqnarray}
	\label{nrep2}	S(a,a^+,b,b^+)=a_l\!^+(\Delta^{-1}{}_{ll'}b_{l'}+b_l\!^+)-ga_l\!^+\Delta^{-1}{}_{ll'}(a_{l'}{}^2-a_{l'}b_{l'}\!^+ -z_{l'}b_{l'}\!^+) .
\end{eqnarray}
 
 This representation can be looked at as a solution of the linear relations (\ref{hs}) of Halpern and Schwartz.
In section 2 a new derivation of these relations will be given, starting directly from the nonlinear Schwinger Dyson equations for the large N theory, without the use of the generalised algebra of Halpern and Schwartz. In sections 3 and 4 these relations will be solved, leading to the new representation for the generating functional of the n-point functions. In section 5 I will derive this new representation directly from the Schwinger Dyson equation for the generating functional. Section 6 contains concluding remarks and an outlook for the numerical treatment of the new representation. An appendix gives a short description of a series expansion for the new representation (\ref{nrep1},\ref{nrep2}). 

I will follow closely the notational framework of \cite{halpern99b}. The object of study is the large N limit of an  Euclidean field theory with NxN Hermitean matrix valued fields $M_i$ in a discretised space of d dimensions. The large N sector of this theory is determined by the vacuum expectation values of the trace of products of n fields (n-point functions or Schwinger functions)
\begin{eqnarray}
	z_{i_1,i_2,\ldots,i_n} = \lim_{N\to\infty}N^{-1}<\!\!v|Tr(M_{i_1}M_{i_2}\cdots M_{i_n})|v\!\!>,
\end{eqnarray}
where $|v\!\!>$ is the vacuum state of the field theory. By definition, $z_{i_1,i_2,\ldots,i_n}$ is invariant under cyclic permutations of its indices, a property which will be used implicitely in the following sections.

The Schwinger Dyson equations for the large N Schwinger functions are nonlinear. For the case of a cubic interaction, which will be considered as an example throughout the following, they read
\begin{eqnarray}
\label{SD}	\Delta_{ij}z_{j,i_1,\ldots,i_n}+gz_{i,i,i_1,\ldots,i_n}=\sum_{k=0}^n\delta_{ii_k}z_{i_1,\ldots,i_{k-1}}     z_{i_{k+1},\ldots,i_n},
\end{eqnarray}
where $\Delta_{ij}$ includes the discretised d-dimensional Laplace operator and a mass term. Repeated incides are summed over the d-dimensional lattice points.

The generating functional for the Schwinger functions has to be defined in terms of non-commuting sources because the Schwinger functions are not symmetric in their arguments. Non-commuting sources were introduced in \cite{cvitanovic82} and also used in \cite{halpern99b}. I will use as non-commuting sources the creation operators $a_i{}\!^+$ and define a generating functional $Z(a^+)$ for n-point functions as
\begin{eqnarray}
\label{fez}
	Z(a^+) & = & \sum_{n=0}^{\infty} \sum_{i_1,\ldots,i_n} a_{i_n}{}\!\!\!^+ \cdots \:a_{i_1}{}\!\!\!^+ z_{i_1,\ldots,i_n}\nonumber\\& = & \sum_{\omega} a^{+\tilde{\omega}} z_{\omega},
\end{eqnarray}
where $\omega$ represents the word of indices $(i_1,\ldots,i_n)$, $\tilde{\omega}$ the word with the elements of $\omega$ in reversed order, and $a^{+\tilde{\omega}}$ denotes the monomial $a_{i_n}{}\!\!\!^+ \cdots \:a_{i_1}{}\!\!\!^+$.

The individual Schwinger functions are recovered by taking the vacuum expectation value of  $Z(a^+)$ multiplied with the appropriate number of annihilation operators :
\begin{eqnarray}
	z_\omega = \bra{a} a^\omega Z(a^+)\ket{a}.
\end{eqnarray}

\section{Linearising the Schwinger Dyson Equations}

As shown in \cite{haan80}, the Schwinger Dyson equations are solved (in the notation of \cite{halpern99b}) by Hermitean master fields $\phi_l$ and Hermitean master momenta $\tilde{\pi}_l$ obeying the "planar" commutation relation
\begin{eqnarray}
  \label{commutation}
	[\phi_l,\tilde{\pi}_{l'}] = i\delta_{ll'}\kett\braa,
\end{eqnarray}
and the vacuum equation
\begin{eqnarray}
\label{vacuum}
	\tilde{\pi}_{l}\kett = i/2 G_l(\phi)\kett,
\end{eqnarray}
where $G_l$ is determined by the Lagrangean of the underlying field theory. For the case of cubic self interaction 
\begin{eqnarray}
	G_l(\phi) = \Delta_{ll'} \phi_{l'} + g\phi_l{}^2.
\end{eqnarray}
The Schwinger Dyson equations (\ref{SD}) follow by equating the evaluation of the expression
\begin{eqnarray}
	\braa [\phi_{i_1}\cdots\phi_{i_n},\tilde{\pi}_{i}]\kett
\end{eqnarray}
using the commutation relation (\ref{commutation}) with the evaluation using the vacuum equation (\ref{vacuum}).

Motivated by the construction of polynomials in the fields with vanishing vacuum expectation values in \cite{halpern99b} (cf. eq. (\ref{hs})), I look for an expression of the form $D(\phi,\beta)=(1-R(\phi,\beta))^{-1}$, depending on an external non-commuting source $\beta_l$, which will be determined by the requirement, that $\braa D(\phi,\beta)\kett=1$. Evaluating the vacuum expectation value of the commutator of $D$ with $\tilde{\pi}_l$ and equating it with the corresponding expression using the vacuum equation results in 
\begin{eqnarray}
\label{e1}
	i\braa G_l(\phi)D(\phi,\beta) \kett = \braa D(\phi,\beta)\:[R(\phi,\beta), \tilde{\pi}_l]\:D(\phi,\beta)\kett .
\end{eqnarray}
Now $R(\phi,\beta)$ should contain a term involving $G_l(\phi)$ in order to cancel the $G_l(\phi)$ on the left hand side. This in turn implies terms proportional to $\phi$, coming from the commutator on the right hand side, which can be canceled only if $R$ contains in addition a term linear in $\phi$. This leads to the Ansatz
\begin{eqnarray}
	R(\phi,\beta) = \beta_l G_l(\phi)+r_l(\beta)\phi_l + r(\beta).
\end{eqnarray}

With this Ansatz the eq. (\ref{e1}) becomes
\begin{eqnarray}	\braa G_l(\phi)D\kett&=&\braa D\kett (\beta_{l'}\Delta_{l'l}+r_l(\beta))\braa D \kett \nonumber\\&&\mbox{}+g\braa D\phi_l\kett \beta_l\braa D \kett \\&&\mbox{}+g\braa D\kett \beta_l\braa \phi_lD\kett. \nonumber
\end{eqnarray}
Multiplyng by $\beta_l$, summing over $l$ and eliminating the factor $\beta_lG_l(\phi)$ from the left hand side yields
\begin{eqnarray}
	D &=& 1 \nonumber\\&&\mbox{}+ (r + \beta_lD(\beta_{l'}\Delta_{l'l}+r_l)+g\beta_lD_l\beta_l)D \nonumber\\
&&\mbox{}+(r_l+g\beta_lD\beta_l)D_l,	
\end{eqnarray}
where $D$ is shorthand for $\braa D \kett$ and $D_l$ for $\braa D\phi_l \kett=\braa \phi_lD \kett$ and both are functions of the noncommuting $\beta$'s.

Repeating the same procedure for the commutator of $D\phi_k$ with $\tilde{\pi}_l$ leads to a similar equation for $D_k$:
\begin{eqnarray}
		D_k &=& z_k+\beta_kD \nonumber\\&&\mbox{}+ (r + \beta_lD(\beta_{l'}\Delta_{l'l}+r_l)+g\beta_lD_l\beta_l)D_k \nonumber\\
&&\mbox{}+(r_l+g\beta_lD\beta_l)D_{lk}.
\end{eqnarray}
Here $z_k=\braa \phi_k \kett$ is the one-point Schwinger function of the theory and $D_{lk}=\braa \phi_lD\phi_k \kett$.
From these equations it is evident, that choosing
\begin{eqnarray}
	r_l(\beta)&=&-g\beta_lD\beta_l,\\
	r &=& -\beta_lD\beta_{l'}\Delta_{l'l} - g(\beta_lD_l\beta_l -\beta_lD\beta_lD\beta_l)
\end{eqnarray}
gives $D=1$ and $D_k=z_k+\beta_k$, which in turn determines $R(\phi,\beta)$ as
\begin{eqnarray}
	R(\phi,\beta)=\beta_lG_l(\phi)-\beta_l\beta_{l'}(\Delta_{l'l}+\delta_{l'l}g(\phi_l+\braa \phi_l \kett)).
\end{eqnarray}
Since $1=D=\braa (1-R(\phi,\beta))^{-1} \kett$, this is a new derivation of the result (\ref{hs}) of Halpern and Schwartz. 

Using the identity
\begin{eqnarray}
	z_\omega=\braa \phi^\omega \kett = \bra{a} a^\omega Z(a^+) \ket{a},
\end{eqnarray}
the result of this section can be reformulated in terms of $Z(a^+)$, the generating funtional of Schwinger functions:
\begin{eqnarray}
\label{lin}
	\bra{a}(1-R(a,\beta))^{-1}Z(a^+)\ket{a} = 1.
\end{eqnarray}

In the next sections, I will invert relation (\ref{lin}) and obtain an explicit expression for $Z(a^+)$ in terms of a second set of free-algebraic operators.

\section{Solution for the Non-Interacting Theory}
In order to solve eq. (\ref{lin}) for $Z(a^+)$ I make the following Ansatz:
\begin{eqnarray}
\label{ansatz}
	(1-R(a,\beta))^{-1}Z(a^+)=1+H(a,\beta)+K(a^+,\beta),
\end{eqnarray}
 where  $H(a,\beta)$ is a sum of products of annihilation operators
\begin{eqnarray}
	H(a,\beta)=\sum_{\left|\omega\right|\ge1}h_{\omega}(\beta)a^{\tilde{\omega}},
\end{eqnarray}
 and $K(a^+,\beta)$ is a sum of products of creation operators 
\begin{eqnarray}
\label{Kexp}
	K(a^+,\beta)=\sum_{\left|\omega\right|\ge1}k_{\omega}(\beta)a^{+{\tilde{\omega}}}.
\end{eqnarray}
 The summation index $\left|\omega\right|\ge1$ indicates that the sum extends over words of one and more letters, such that $\bra{a}(H(a,\beta)\ket{a}=0$ and $\bra{a}(K(a^+,\beta)\ket{a}=0$ and therefore the Ansatz fullfills  eq. (\ref{lin}). Terms containing mixed products of creation and annihilation operators cannot appear, because the left hand side is a product of an operator containing only annihilation operators and an operator containing only creation operators and all mixed products of $a$'s and $a^+$'s can be eliminated using (\ref{prod}).
 
Multiplying (\ref{ansatz}) by $1-R(a,\beta)$ and collecting terms containing annihilation and creation operators results in separate equations for $H$ and $K$. In order to explain the solution method, I will first consider the  non-interacting case with 
\begin{eqnarray}
	R\ze(a,\beta) = \Delta_{ll'}(a_l\beta_{l'}-\beta_l\beta_{l'}).
\end{eqnarray}
From this the equations for $H$ and $K$ follow:
\begin{eqnarray}
	(1-R\ze(a,\beta))H(a,\beta)&=&\Delta_{ll'}a_l\beta_{l'}\\
	\label{Keq}
	(1+\Delta_{ll'}\beta_l\beta_{l'})(1+K(a^+,\beta))-\Delta_{ll'}a_l\beta_{l'}K(a^+,\beta)&=&Z\ze(a^+).
\end{eqnarray}
The annihilation part will not be needed further on. The equation for $K$ leads to the following iterative equations for $k_{\omega}(\beta)$:
\begin{eqnarray}
\label{it}
	\Delta_{ll'}\beta_lk_{\omega l'}(\beta)=(1+\Delta_{ll'}\beta_l\beta_{l'})k_\omega(\beta) - k_\omega(0),
\end{eqnarray}
where $z\ze_{\omega}=k_{\omega}(0)$ has been used. The iteration can be made explicit by separating the constant term in $k_\omega$ as $k_\omega(\beta)=k_\omega(0)+\beta_lk^{(l)}_\omega(\beta)$:
\begin{eqnarray}
\label{it1}
	k_{\omega l}(\beta)=\Delta^{-1}{}_{ll'}k^{(l')}_\omega(\beta)+\beta_lk_\omega(\beta).
\end{eqnarray}

Up to now, the non-commuting sources $\beta_l$ have not been explicitly specified. By choosing for the non-commuting sources creation operators $b^+\!\!{}_l$ in a free Fock space with vacuum $\ket{b}$, the solution will become an explicit expression in terms of operators in this Fock space. In terms of these creation operators, eq. (\ref{Keq})  reads
\begin{eqnarray}
\label{Keqn}
		(1+\Delta_{ll'}b_l{}\!^+b_{l'}{}\!^+)(1+K(a^+,b^+))-\Delta_{ll'}a_lb_{l'}{}\!^+K(a^+,b^+)&=&Z\ze(a^+).
\end{eqnarray}
 
The iterative equations (\ref{it1}) now can be combined into a single equation obtained form (\ref{Keqn}) by multiplication from the left with the annihilating operator $b_l$:
\begin{eqnarray}
\label{itn}
	a_lK(a^+,b^+) = (\Delta^{-1}{}_{ll'}b_{l'}+b_l{}\!^+)(1+K(a^+,b^+)-\Delta^{-1}{}_{ll'}b_{l'}Z\ze(a^+)
\end{eqnarray}
Since $K(a^+,b^+)$ has no term independent of $a^+$, $K(a^+,b^+)=a^+\!\!{}_la_lK(a^+,b^+)$. Using this identity in the expression for $1+K(a^+,b^+)$ and inserting eq. (\ref{itn}) for $a_lK(a^+,b^+)$ leads to the following solution for $1+K(a^+,b^+)$:
\begin{eqnarray}
\label{Keqn1}
	1+K(a^+,b^+)=[1-a_l{}\!^+(\Delta^{-1}{}_{ll'}b_{l'}+b_l{}\!^+)]^{-1}(1-\Delta^{-1}{}_{ll'}b_{l'}a_l{}\!^+Z\ze(a^+)).
\end{eqnarray}

According to (\ref{Keqn}), $Z\ze(a^+)=\bra{b}(1+K(a^+,b^+))\ket{b}$. Taking the vacuum expectation value of (\ref{Keqn1}) eliminates  the term containing $Z\ze(a^+)$ from the right hand side and therefore gives the explicit form for $Z\ze(a^+)$ in terms of the free-algebraic operators $b_l$ and $b_l{}\!^+$:
\begin{eqnarray}
\label{freezeq}
	Z\ze(a^+)=\bra{b}[1-a_l{}\!^+(\Delta^{-1}{}_{ll'}b_{l'}+b_l{}\!^+)]^{-1}\ket{b}.
\end{eqnarray}

\section{Solution for the Interacting Theory}

For the interacting theory similar methods as in the last section can be applied. For the cubic self interaction the generating functional for polynomials with vanishing vacuum expectation values is built from the following $R$:
\begin{eqnarray}
	R(a,b^+)=b_l\!^+(\Delta_{ll'}a_{l'}+ga_{l}{}\!^2) -b_l{}\!^+b_{l'}{}\!^+(\Delta_{ll'}+g\delta_{ll'}(a_l+z_l)).
\end{eqnarray}
 Analogously to (\ref{Keqn}) the following equation for $K$  can be derived in the case of cubic interaction: 
\begin{eqnarray}
\label{int}
	(1+\Delta_{ll'}b_l{}\!^+b_{l'}{}\!^++gz_lb_l{}\!^+{}^2)(1+K(a^+,b^+))&&\nonumber\\ \mbox{}-(b_{l}{}\!^+\Delta_{ll'}a_{l'} -g b_{l}{}\!^+{}^2a_l) K(a^+,b^+)&&\nonumber\\ \mbox{}-gb_l{}\!^+a_l{}\!^2(K(a^+,b^+)-a_l{}\!^+k_l(b^+))&&\hspace{-1.5em}=Z(a^+).
\end{eqnarray}
In (\ref{int}) all terms containing zero and more creation operators $a^+$ are collected from the equation $(1-R)(1+K+H)= Z$. Compared to the non-interacting case (\ref{Keqn}) the third line on the left hand side of (\ref{int}) contains an additional factor with two annihilation operators multiplying $K$. The term linear in $a^+$ in the expansion of K therefore would contribute an annihilation operator and has to be subtracted from $K$.  

The same steps as in the non-interacting case now lead to the following equation for $1+K$:
\begin{eqnarray}
\label{iKeq}	1+K(a^+,b^+)=\nonumber\\(1+ga_l{}\!^+\Delta^{-1}{}_{ll'}(a_{l'}{}\!^2-a_{l'}b_{l'}{}\!^+ -z_{l'}b_{l'}{}\!^+)-a_l{}\!^+(\Delta^{-1}{}_{ll'}b_{l'}+b_l{}\!^+)^{-1}\\\nonumber(1+ga_l{}\!^+\Delta^{-1}{}_{ll'}a_{l'}(k_{l'}(b^+)-b_{l'}{}\!^+ +a_{l'})-a_l{}\!^+\Delta^{-1}{}_{ll'}b_{l'}Z(a^+)).
\end{eqnarray}
The right hand side of this equation still contains the unknown functions $Z(a^+)=\bra{b}(1+K(a^+,b^+)\ket{b}$ and $k_l(b^+)=\bra{a}aK(a^+,b^+)\ket{a}$, but these can be determined from (\ref{iKeq}) by taking the appropriate vacuum expectation values. With the abbreviation 
\begin{eqnarray}	S(a,a^+,b,b^+)=a_l\!^+(\Delta^{-1}{}_{ll'}b_{l'}+b_l{}\!^+)-ga_l\!^+\Delta^{-1}{}_{ll'}(a_{l'}{}\!^2-a_{l'}b_{l'}{}\!^+ -z_{l'}b_{l'}{}\!^+)
\end{eqnarray}
in the denominator in the r.h.s. of (\ref{iKeq}) this leads to
\begin{eqnarray}
\label{f1}
	Z(a^+)&=&\bra{b}(1-S)^{-1}(1+ga_l{}\!^+\Delta^{-1}{}_{ll'}a_{l'}(k_{l'}(b^+)-b_{l'}{}\!^+ + a_{l'}))\ket{b}\\
\label{f2}
	k_i(b^+)&=&\bra{a}a_i(1-S)^{-1}(1-a_l{}\!^+\Delta^{-1}{}_{ll'}b_{l'}Z(a^+)\ket{a}. 
\end{eqnarray}

When the expression (\ref{f2}) for $k_l$ is inserted into (\ref{f1}), the term proportional to $Z(a^+)$ in (\ref{f2}) does not contribute, because it contains the factor $b_{l'}$, which annihilates the vacuum in eq.(\ref{f1}). Eq. (\ref{f1}) therefore gives an explicit solution for $Z(a^+)$. 

Of course for the Schwinger functions $z_\omega$ the additional terms in (\ref{f1}) do not contribute, because the annihilation operators $a_l$ act directly on the vacuum $\ket{a}$:
\begin{eqnarray}
\label{xxx}
	z_\omega=\bra{a}\bra{b}a^\omega(1-S(a,a^+,b,b^+))^{-1}\ket{b}\ket{a}.
\end{eqnarray}

This is the central result of the present work, expressing the large $N$ Schwinger functions as vacuum expectation values of an operator constructed from two sets of free-algebraic operators. The expression (\ref{xxx}) still is not fully explicit, because the operator $S(a,a^+,b,b^+)$  contains the one-point function $z_l$, which has to be determined selfconsistently from this expression with $\omega=l$.

\section{Direct Solution from the Schwinger Dyson \\ Equations}

With the insight from the last sections the Schwinger Dyson equations can be solved directly, leading to the same solution (\ref{xxx}) for the planar Schwinger functions. Multiplying the Schwinger Dyson equations in the form
\begin{eqnarray}
	z_{i\omega}=\Delta^{-1}{}_{ij}(\sum_{\omega=\omega_1j\omega_2}z_{\omega_1}z_{\omega_2} - g z_{jj\omega})
\end{eqnarray}
by $a^+{}^{\tilde{\omega}}$ and summing over $\omega$ gives
\begin{eqnarray}
\label{sd}
	a_i(Z(a^+)-1) = \Delta^{-1}{}_{ij}(Z(a^+)a_j{}\!^+Z(a^+)-g(a_j{}\!^2Z(a^+)-a_j{}\!^2 - z_ja_j).
\end{eqnarray}
In this derivation the following identities have been used:
\begin{eqnarray}
	 &&\sum_{\omega}a^+{}^{\tilde{\omega}} z_{i\omega}=\sum_{\omega}a^+{}^{\tilde{\omega}} z_{\omega i}=a_i(\sum_{\omega}a^+{}^{\tilde{\omega}} z_{\omega}-1) = a_i(Z(a^+)-1)\\
	 && \sum_{\omega}a^+{}^{\tilde{\omega}} z_{jj\omega}=\sum_{\omega}a^+{}^{\tilde{\omega}} z_{\omega jj} \\\nonumber &&=a_j{}\!^2(\sum_{\omega}a^+{}^{\tilde{\omega}} z_{\omega}-1-a_l{}\!^+z_l) = a_j{}\!^2(Z(a^+)-1-a_l{}\!^+z_l).
\end{eqnarray}
Inserting (\ref{sd}) into the identity $Z(a^+)=1 + a_i{}\!^+a_i(Z(a^+)-1)$ leads to the following nonlinear functional equation for $Z(a^+)$:
\begin{eqnarray}
\label{nonlZ}	Z(a^+)=(1+ga_i{}\!^+\Delta^{-1}{}_{ij}a_j{}\!^2-a_i{}\!^+\Delta^{-1}{}_{ij}Z(a^+)a_j{}\!^+)^{-1}\nonumber\\(1+ga_i{}\!^+\Delta^{-1}{}_{ij}(a_j{}\!^2+a_jz_j)).
\end{eqnarray}

The corresponding functional equation for the non-interacting case ($g=0$) has been given in \cite{cvitanovic82}:
\begin{eqnarray}
\label{nonl}	
Z\ze(a^+)=(1 -a_i{}\!^+\Delta^{-1}{}_{ij}Z^{(0)}(a^+)a_j{}\!^+)^{-1}.  
\end{eqnarray}
The solution for $Z\ze(a^+)$ can be constructed using the explicit realisation of the non-interacting planar field from \cite{haan80} in terms of free-algebraic operators , $\phi\ze_i=\Delta^{-1/2}{}_{ij}(b_j+b_j{}\!^+)$:
\begin{eqnarray}
\label{expl}	Z\ze(a^+)=\sum_{\omega}a^+{}^{\tilde{\omega}}<\!\!0|\phi\ze{}^\omega|0\!\!>=\bra{b}(1-a_l{}\!^+\Delta^{-1/2}{}_{ij}(b_j+b_j{}\!^+))^{-1}\ket{b}.
\end{eqnarray}
Here it has been used, that because $<\!\!0|\phi\ze{}^\omega|0\!\!>$ is real it equals $<\!\!0|\phi\ze{}^{\tilde{\omega}}|0\!\!>$.
The result (\ref{expl}) should be compared to the solution for $Z\ze(a^+)$ found in sect. 3 from the linearised form of the Schwinger Dyson equations. The difference in powers of the propagator $\Delta_{ij}$ is irrelevant as will be demonstrated below.  

It will be illuminating to prove explicitly, that the representation (\ref{expl}) is a solution to the nonlinear equation (\ref{nonl}). I start with a more general expression
\begin{eqnarray}
\label{repF}
	F(\lambda,\rho)= \bra{b}(1-\lambda_jb_j-\rho_jb_j{}\!^+))^{-1}\ket{b},
\end{eqnarray}
where $\lambda_l$ and $\rho_l$ are objects commuting with $b_l$ and $b_l{}\!^+$ but not necessarily with each other.
Expanding the denominator gives $F$ as a sum of expectation values $F_n$ of powers :
\begin{eqnarray}
F_n=\bra{b}(\lambda_jb_j+\rho_jb_j{}\!^+)^n\ket{b}.
\end{eqnarray}
Each term $F_n$ is the sum of all possible ways to contract pairs of $b_l$ and $b_l{}\!^+$ according to the rule $b_lb_{l'}{}\!^+=\delta_{ll'}$ until no operators $b_l$ or $b_l{}\!^+$ are left over. This leads to the following recursion for $F_n$: From the leftmost factor $\lambda_jb_j+\rho_jb_j{}\!^+$ the creation operator $b_l{}\!^+$ annihilates $\bra{b}$. The annihilation operator $b_l$ can be contracted with a creation operator in any of the remaining $n-1$ factors. A contraction with the k-th factor implies, that all $b$'s and $b^+$'s of the factors 2 to $k-1$ have been contracted completely among themselves to yield $F_{k-2}$, and similarly the factors $k+1$ to $n$ must give $F_{k-n}$. Therefore
\begin{eqnarray}
	F_n=\sum_{k=2}^n\lambda_jF_{k-2}\rho_jF_{n-k}.
\end{eqnarray}
Resumming the $F_n$ and using $F_0=1$ and $F_1=0$ yields the equations 
\begin{eqnarray}
	F=1+\lambda_lF\rho_lF
\end{eqnarray}
and 
\begin{eqnarray}
\label{nonlF}
	F=(1-\lambda_lF\rho_l)^{-1}.
\end{eqnarray}

With either
$\lambda_l=\rho_l=a_{l'}{}^+\Delta^{-1/2}{}_{l'l}$ or $\lambda_l=a_{l'}{}^+\Delta^{-1}{}_{l'l}$, $\rho_l=a_{l}{}^+$, $F$ obeys the defining equation (\ref{nonl}) for the generating functional of the non-interacting theory. This also shows, that the expressions for $Z\ze(a^+)$ given in (\ref{expl}) and in (\ref{freezeq}) of sect. 3 are equivalent.

The result (\ref{nonlF}) for $F(\lambda,\rho)$ now also can  be used to obtain a representation for $Z(a^+)$ in the interacting case. This becomes apparent, if eq. (\ref{nonlZ}) is rewritten as
\begin{eqnarray}
\label{nonls}
	Z=(1+A -L_jZR_j)^{-1}(1+B),
\end{eqnarray}
with obvious definitions for $A$, $B$, $L_j$ and $R_j$, and if Z is transformed to
\begin{eqnarray}
\label{tran}
	F=Z(1+B)^{-1}(1+A).
\end{eqnarray}
Rewriting (\ref{nonls}) as an equation for $F$ reproduces the nonlinear equation (\ref{nonlF}) with
\begin{eqnarray}
	\lambda_j=(1+A)^{-1}L_j, \;\; \rho_j=(1+A)^{-1}(1+B)R_j.
\end{eqnarray}
 
Inverting the transformation (\ref{tran}) and inserting the representation (\ref{repF}) for $F$ now yields the explicit solution for $Z(a^+)$ in terms of the auxiliary operators $b$ and $b^+$:
\begin{eqnarray}
\label{sdf1}	Z(a^+)=&&\hspace{-1.5em}\bra{b}(1+ga_l{}^+\Delta^{-1}{}_{ll'}(a_{l'}{}\!^2-a_{l'}b_{l'}{}\!^+-z_{l'}b_{l'}{}\!^+)-a_l{}\!^+(\Delta^{-1}{}_{ll'}b_{l'}+b_{l'}{}\!^+))^{-1}\nonumber\\&&\hspace{-1.5em}(1+ga_l{}\!^+\Delta^{-1}{}_{ll'}(a_{l'}{}\!^2+z_{l'}a_{l'}))\ket{b}.
\end{eqnarray}

Comparing this expression for $Z$ with eq. (\ref{f1}) from sect.~4, there is a difference in the additional terms multiplying the denominator: instead of  $ga_l{}\!^+\Delta^{-1}{}_{ll'}(k_{l'}(b^+)-b_{l'}{}\!^+)a_{l'}$ in (\ref{f1}) here appears the term  $ga_l{}^+\Delta^{-1}{}_{ll'}z_{l'}a_{l'}$. Of course the additional terms multiplying the denominator do not contribute when evaluating the expectation values $z_\omega=\bra{a}a^\omega Z(a^+)\ket{a}$, because they annihilate $\ket{a}$. Because the Schwinger functions $z_\omega$ completely determine $Z(a^+)$ (cf. (\ref{fez}), the expressions (\ref{f1}) and (\ref{sdf1}) must coincide. In the apppendix, this  will be shown explicitly up to third order in the interaction parameter $g$.

\section{Conclusion and Outlook}
In this work an explicit representation of the generating functional for the large N limit of Schwinger functions has been found in terms of free-algebraic operators. It is to be hoped, that this representation can be used for the numerical evaluation of the large N limit of field theories. Perhaps in the future methods  for the representation of free-algebraic operators will be developed, which allow for a non-perturbative treatment of the planar Schwinger functions. 

For the present I see the possibility to use the new expression for the large N Schwinger functions,
\begin{eqnarray}
\label{rep}	\lefteqn{z_\omega=}\hspace{-12em}\\\nonumber&&\hspace{-2em}\bra{a}\bra{b}a^\omega(1+ga_l{}\!^+\Delta^{-1}{}_{ll'}(a_{l'}{}\!^2-a_{l'}b_{l'}{}\!^+-z_{l'}b_{l'}{}\!^+)-a_l{}\!^+(\Delta^{-1}{}_{ll'}b_{l'}+b_{l'}{}\!^+))^{-1}\ket{b}\ket{a}
\end{eqnarray}
as the starting point for a perturbative expansion. 

An expansion of the denominator yields a sum of graphs with two kinds of links for the contraction of $a$- and $b$-operators. The use of two kinds of links in the description of large N perturbation series is not new. The very first paper discussing the number of planar Feynman diagrams \cite{koplik77}
separated the links of planar graphs into two classes in order to deduce an upper limit for their number. In effect \cite{koplik77} already contains the proof, that for the zero dimensional case the representation (\ref{repF}) leads to the nonliner equation (\ref{nonlF}). Also in a more recent series of papers (\cite{schaeffer97}, \cite{bouttier02} and references therein) a combinatorial construction of the planar limit in zero dimensions was given which uses two kinds of links.  

The explicit representation (\ref{rep}) furnishes a very simple set of rules for the generation of planar graphs. These rules are summarised in the appendix. The use of these rules in a Monte Carlo procedure for the evaluation of the sum of planar diagrams will be developed in future work.

\section*{Appendix}

In this appendix an expansion of the denominator in the representation (\ref{rep}) will be developed. Using this expansion, the equivalence of the formulae (\ref{f1}) and (\ref{sdf1}) for $Z(a^+)$ will be proven up to third order in the coupling constant $g$.   

The denominator $1- S(a,a^+,b,b^+)$ in the representation of planar Schwinger functions
involves two sets of free-algebraic operators $a_l$, $a_l{}\!^+$ and $b_l$, $b_l{}\!^+$ in five different combinations:
 \begin{eqnarray*}
	{\cal O}^{(1)}&=& a_{l}{}^+b_l{}\!^+  \\
	{\cal O}^{(2)}&=& a_{l}{}\!^+\Delta^{-1}{}_{ll'}b_l\\
	{\cal O}^{(3)}&=& -ga_{l}{}\!^+\Delta^{-1}{}_{ll'}a_{l'}{}\!^2 \\
	{\cal O}^{(4)}&=& ga_{l}{}\!^+\Delta^{-1}{}_{ll'}z_{l'}b_{l'}{}\!^+ \\
	{\cal O}^{(5)}&=& ga_{l}{}\!^+\Delta^{-1}{}_{ll'}a_{l'}b_{l'}{}\!^+.
\end{eqnarray*}

Expanding the denominator yields a geometric series of products of operators
\begin{eqnarray}
\label{sum}
	(1-S(a,a^+,b,b^+)^{-1}=1+\sum_{n=1}^\infty \hspace{0.5em}\sum_{i_1,\cdots,i_n = 1}^5 {\cal O}^{(i_1)} \cdots {\cal O}^{(i_n)}
\end{eqnarray}
and the action of every term of this series on the combined vacuum state $\ket{b}\ket{a}$ can be interpreted as a walk in the two-dimensional $(n_a,n_b)$-plane of states occupied by $n_a$ particles of type $a$ and by $n_b$ particles of type $b$. In this plane the state $\ket{b}\ket{a}$ is represented by the origin and each of the five operators ${\cal O}^{(i)}$ can be represented as a specific arrow  \boldmath $a^{(i)}$ \unboldmath indicating the shift of occupation numbers induced by the action of the operator:
\vspace{2em}

\unitlength0.4cm
\begin{picture}(16,1)
\put(0,0){${\cal O}^{(1)} \Longrightarrow$ \hspace{3em}\boldmath $a^{(1)} =$ \unboldmath }
\put(10,0){\vector(1,1){1}}
\end{picture}
\vspace{1em}

\begin{picture}(6,1)
\put(0,0){${\cal O}^{(2)} \Longrightarrow$ \hspace{3em}\boldmath $a^{(2)} =$ \unboldmath }
\put(10,1){\vector(1,-1){1}}
\end{picture}
\vspace{1em}

\begin{picture}(6,1)
\put(0,0){${\cal O}^{(3)} \Longrightarrow$ \hspace{3em}\boldmath $a^{(3)} =$ \unboldmath }
\put(11,0){\vector(-1,0){1}}
\end{picture}
\vspace{1em}

\begin{picture}(6,1)
\put(0,0){${\cal O}^{(4)} \Longrightarrow$ \hspace{3em}\boldmath $a^{(4)} =$ \unboldmath }
\put(10,0){\vector(1,1){1}}
\put(10.1,-0.1){\vector(1,1){1}}
\end{picture}
\vspace{1em}

\begin{picture}(6,1)
\put(0,0){${\cal O}^{(5)} \Longrightarrow$ \hspace{3em}\boldmath $a^{(5)} =$ \unboldmath }
\put(10,0){\vector(0,1){1}}
\end{picture}
\vspace{1em}

\begin{figure}[ht]
\hspace*{12em}
\begin{picture}(6,1)
\put(0,0){\vector(1,1){1}}
\put(1,1){\vector(1,-1){1}}
\put(2,0){\vector(0,1){1}}
\put(2,1){\vector(1,1){1}}
\put(2.1,0.9){\vector(1,1){1}}
\put(3,2){\vector(-1,0){1}}
\put(0,0){\vector(1,0){6}}
\put(7,0){$n_a$}
\put(0,0){\vector(0,1){4}}
\put(0,5){$n_b$}
\end{picture}
\caption{\label{example}A walk contributing to the sum (\ref{sum})}
\end{figure}
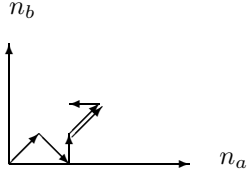

In order to distinguish the operators ${\cal O}^{(1)}$ and ${\cal O}^{(4)}$, both of which increase $n_a$ and $n_b$ by one unit,  ${\cal O}^{(4)}$ is represented by a double line arrow. 
As an example fig. (\ref{example}) shows the path representing the term  ${\cal O}^{(3)}{\cal O}^{(4)}{\cal O}^{(5)}{\cal O}^{(2)}{\cal O}^{(1)}\ket{b}\ket{a}$.  
\vspace{7em}

\vspace{2em}

Not all walks generated by the five arrows contribute to $(1-S)^{-1}\ket{b}\ket{a}$. Since ${\cal O}^{(2)}$ annihilates $\ket{b}$, ${\cal O}^{(4)}$ annihilates $\ket{a}$ and ${\cal O}^{(3)}$ annihilates both $\ket{a}$ and $a_l{}\!^+\ket{a}$ the first step of the walk has to go from $(0,0)$ to $(1,1)$ and all sites visited later on must have $n_a \ge 1$ and $n_b \ge 0$. Furthermore because of the identity ${\cal O}^{(3)}{\cal O}^{(1)}=-{\cal O}^{(5)}$, every contribution to the sum containing the operator ${\cal O}^{(5)}$ will be canceled by a corresponding contribution containing the product ${\cal O}^{(3)}{\cal O}^{(1)}$ instead of ${\cal O}^{(5)}$. This can be summarised by the following set of rules for the generation of walks representing the sum (\ref{sum}):
\begin{itemize}
\item[r1 - ] All walks are constructed from the four arrows \boldmath $a^{(1)},a^{(2)},a^{(3)},a^{(4)}$ \unboldmath %\vspace{1em} \\
%\begin{picture}(2,1)
%\put(0,0){\vector(1,1){1}}
%\end{picture}
%\begin{picture}(2,1)
%\put(0,1){\vector(1,-1){1}}
%\end{picture}
%\begin{picture}(2,1)
%\put(1,0){\vector(-1,0){1}}
%\end{picture}
%\begin{picture}(2,1)
%\put(0,0){\vector(1,1){1}}
%\put(0.1,-0.1){\vector(1,1){1}}
%\end{picture}
\item[r2 - ] The first step of the walk must be \boldmath $a^{(1)}$ \unboldmath or \boldmath $a^{(4)}$ \unboldmath
\item[r3 - ] The walk must not extend to the left of $n_a=1$ or below $n_b=0$
\item[r4 - ] \boldmath $a^{(1)}$ \unboldmath must never be followed by \boldmath $a^{(3)}$ \unboldmath
\end{itemize}

The one-point planar Schwinger function $z_l$ is the sum of all walks starting at the point $(0,0)$ and ending at $(1,0)$. To zeroth order in the coupling $g$ only walks built up from the operators ${\cal O}^{(1)}$ and ${\cal O}^{(2)}$ contribute, because all other operators include a factor of $g$. Since it it not possible to construct a walk
from $(0,0)$ to $(1,0)$ using only \boldmath $a^{(1)}$ \unboldmath and \boldmath $a^{(2)}$ \unboldmath, there is no zeroth order contribution to $z_l$. 

\begin{figure}[ht]
\hspace*{12em}
\begin{picture}(6,1)
\thicklines
\put(0,0){\vector(1,1){1}}
\put(1,1){\vector(1,-1){1}}
\put(2,0){\vector(-1,0){1}}
\thinlines
%\linethickness{0.05mm}
\put(0,-0.1){\vector(1,0){6}}
\put(7,0){$k_a$}
\put(0,-0.1){\vector(0,1){4}}
\put(0,5){$k_b$}
\end{picture}
\vspace{2em}
\caption{\label{order1}Walk for 1th order $z_l$}
\end{figure}
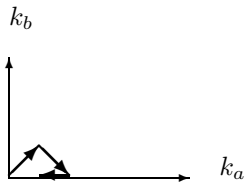

To first order all walks with one operator ${\cal O}^{(3)}$ and any number of operators ${\cal O}^{(1)}$ and ${\cal O}^{(2)}$ contribute. Operator ${\cal O}^{(4)}$ starts at $g^2$, so it cannot be used for first order walks. The only possible walk with this composition is drawn in fig. (\ref{order1}).

\vspace{7em}

The evaluation of the corresponding vacuum expectation value of the operator product yields
\begin{eqnarray}
	z_l{}^{(1)}=-g\Delta^{-1}{}_{ll'}\Delta^{-1}{}_{l'l'}
\end{eqnarray}

To second order, to which products of either two ${\cal O}^{(3)}$ or one ${\cal O}^{(4)}$ and any number of ${\cal O}^{(1)}$ and ${\cal O}^{(2)}$ operators contribute, again no walk connecting $(0,0)$ to $(1,0)$ can be constructed.

\begin{figure}[ht]
\hspace*{1em}
\begin{picture}(5,1)
\thicklines
\put(0,0){\vector(1,1){1}}
\put(1,1){\vector(1,1){1}}
\put(2,2){\vector(1,-1){1}}
\put(3,1){\vector(1,-1){1}}
\put(4,0){\vector(-1,0){1}}
\put(3,0){\vector(-1,0){1}}
\put(2,0){\vector(-1,0){1}}
\end{picture}
\begin{picture}(5,1)
\thicklines
\put(0,0){\vector(1,1){1}}
\put(1,1){\vector(1,1){1}}
\put(2,2){\vector(1,-1){1}}
\put(3,1){\vector(-1,0){1}}
\put(2,1){\vector(1,-1){1}}
\put(3,0){\vector(-1,0){1}}
\put(2,0){\vector(-1,0){1}}
\end{picture}
\begin{picture}(5,1)
\thicklines
\put(0,0){\vector(1,1){1}}
\put(1,1){\vector(1,1){1}}
\put(2,2){\vector(1,-1){1}}
\put(3,1){\vector(-1,0){1}}
\put(2,1){\vector(-1,0){1}}
\put(1,1){\vector(1,-1){1}}
\put(2,0){\vector(-1,0){1}}
\end{picture}
\begin{picture}(5,1)
\thicklines
\put(0,0){\vector(1,1){1}}
\put(1,1){\vector(1,-1){1}}
\put(2,0){\vector(-1,0){1}}
\put(1,0){\vector(1,1){1}}
\put(2,1){\vector(1,-1){1}}
\put(3,0){\vector(-1,0){1}}
\put(2,0){\vector(-1,0){1}}
\end{picture}
\begin{picture}(5,1)
\thicklines
\put(0,0){\vector(1,1){1}}
\put(0.1,-0.1){\vector(1,1){1}}
\put(1,1){\vector(1,-1){1}}
\put(2,0){\vector(-1,0){1}}
\end{picture}
\vspace{2em}
\caption{\label{order3}Walks for 3rd order $z_l$}
\end{figure}
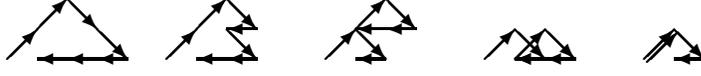

Third order finally receives contributions from 4 walks containing three factors of ${\cal O}^{(3)}$ and from one walk with one ${\cal O}^{(3)}$ and one ${\cal O}^{(4)}$, which are all depicted in fig. (\ref{order3}).
\vspace{7em}

The evaluation of these contributions shows, that the 2nd and 3rd walk have identical values and that the 5th walk cancels one of those which leaves three different contributions to third order:
\begin{eqnarray}	z_l{}^{(3)}=-g^3(&&\hspace{-1.5em}\Delta^{-1}{}_{lk}\Delta^{-1}{}_{km}\Delta^{-1}{}_{kn}\Delta^{-1}{}_{mn}\Delta^{-1}{}_{nm}\nonumber\\&&\hspace{-2.6em}\mbox{}+\Delta^{-1}{}_{lk}\Delta^{-1}{}_{km}\Delta^{-1}{}_{km}\Delta^{-1}{}_{mn}\Delta^{-1}{}_{nn}\\&&\hspace{-2.6em}\mbox{}+\Delta^{-1}{}_{lk}\Delta^{-1}{}_{km}\Delta^{-1}{}_{kn}\Delta^{-1}{}_{mm}\Delta^{-1}{}_{nn}\hspace{0.5em})\nonumber
\end{eqnarray}
 
The factor $k_l(b^+)\ket{b}$, which appears in the expression (\ref{f1}) for $Z(a^+)$ also can be analysed according to the walks, which contribute to the expansion (\ref{sum}) of the denominator in
\begin{eqnarray}
   k_l(b^+)\ket{b}=\bra{a}a_l(1-S)^{-1}\ket{a}\ket{b}.
\end{eqnarray}
All walks according to rules r1 - r4 starting at $(0,0)$ and ending at the line $n_a=1$ will contribute in this expansion. To zeroth order this is just the single step walk \boldmath $a^{(1)}$ \unboldmath, which contributes the term $b_l{}^+\ket{b}$, which cancels the explicit term $-b_l{}^+\ket{b}$ in the expression (\ref{f1}).

To first order all walks must contain one factor of ${\cal O}^{(3)}$, therefore the walk \boldmath$a^{(1)}a^{(2)}a^{(3)}$\unboldmath (cf. fig. \ref{order1}) contributes, whereas the walk \boldmath $a^{(1)}a^{(1)}a^{(3)}$ \unboldmath is forbidden according to rule r4. This shows the equality of $k_l(b^+)\ket{b}$ and $z_l\ket{b}$ to first order in $g$.

\begin{figure}[ht]
\hspace*{12em}
\begin{picture}(5,1)
\thicklines
\put(0,0){\vector(1,1){1}}
\put(1,1){\vector(1,1){1}}
\put(2,2){\vector(1,-1){1}}
\put(3,1){\vector(-1,0){1}}
\put(2,1){\vector(-1,0){1}}
\end{picture}
\begin{picture}(5,1)
\thicklines
\put(0,0){\vector(1,1){1}}
\put(0.1,-0.1){\vector(1,1){1}}
\end{picture}
\vspace{2em}
\caption{\label{korder2}Walks for 2nd order $k_l$}
\end{figure}
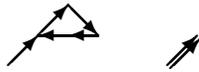

To second order two walks contribute (cf. fig. \ref{korder2}), which cancel each other.
 
\vspace{3em}

The equality of the expression for $Z(a^+)$ obtained in sect. 4, eq.(\ref{f1}), and in sect.5, eq.(\ref{sdf1}), therefore is established up to third order in the coupling constant.

\end{document}